\documentclass[12pt]{article}
\usepackage{cite,epsfig,amssymb,amsmath,graphicx,color}
\evensidemargin \oddsidemargin

\newcommand{\nn}{\nonumber}

\begin{document}

\begin{center}
{{{\Large \bf Janus ABJM Models with Mass Deformation}
}\\[17mm]
Kyung Kiu Kim$^{1}$ and O-Kab Kwon$^{2}$\\[3mm]
{\it $^{1}$Department of Physics, Sejong University, Seoul 05006 Korea\\
$^{2}$Department of Physics,~BK21 Physics Research Division,
~Institute of Basic Science, Sungkyunkwan University, Suwon 440-746, Korea}\\[2mm]
{\it kimkyungkiu@sejong.ac.kr,~okab@skku.edu} }

\end{center}
\vspace{15mm}

\begin{abstract}
We construct a large class of ${\cal N} = 3$ Janus ABJM models with mass deformation, where the mass depends on a spatial (or lightcone) coordinate. We also show that the resulting Janus model can be identified with an effective action of M2-branes in the presence of a background self-dual 4-form field strength varying along one spatial (or lightcone) coordinate. 
\end{abstract}

\newpage
\tableofcontents

\section{Introduction}

The ABJM theory is a 3-dimensional $\mathcal{N} = 6$ Chern-Simon matter theory describing dynamics of M2-branes in M-theory \cite{Aharony:2008ug}. It can be extended to a non-conformal field theory by a relevant deformation, for instance, the mass-deformed ABJM (mABJM) theory~\cite{Hosomichi:2008jb,Gomis:2008vc}. Although the mABJM theory is not  a conformal theory, the deformation still preserves $\mathcal{N} = 6$ supersymmetry and it has a reliable gravity dual, which is known as the 11-dimensional supergravity on the Lin-Lunin-Maldacena (LLM) solutions~\cite{Lin:2004nb}. This gravity dual has been studied in many 
contexts\cite{Kim:2010mr,Cheon:2011gv,Jang:2016tbk}.
Due to the properties of the gauge/gravity 
duality~\cite{Maldacena:1997re,Gubser:1998bc,Witten:1998qj}, one can study the strong coupling limit of the mass-deformed ABJM theory. 
This is one of the advantages to study the mABJM theory, which has the explicit gravity dual.

In addition to the above advantage, the 3-dimensional Chern-Simons theory itself is interesting because it exhibits very different behavior from the other gauge theories. For instance, since there are no propagating degrees of freedom of the gauge field, the Chern-Simons theory encodes the topology of the space. This property can be important in condensed matter applications
and open up the possibility of using mABJM theory.

 There is another interesting deformation of supersymmetric field theories. This deformed theory is called the Janus field theory and its gravity dual is termed the Janus solution. This deformation has a clear physical meaning and provides interesting examples in the gauge/gravity duality. A Janus field theory can be interpreted as a theory with an interface, which divides two spatial regions. In each spatial region, the coupling constant has a different value. This configuration is naturally realized by a Janus solution with a nontrivial dilaton profile through the gauge/gravity duality. Janus solutions were studied mainly in type IIB supergravity and related gauged supergravity theories \cite{Bak:2003jk,Clark:2005te,DHoker:2006vfr,Suh:2011xc}. The Janus deformation of the $\mathcal N$ = 4 Super Yang-Mills (SYM) theory is classified according to supersymmetry. See \cite{Clark:2004sb,DHoker:2006qeo,Kim:2008dj,Choi:2017kxf, Gaiotto:2008sd} for the explicit construction. This interface can be extended to a nullike interface \cite{Chu:2006pa}.

Since there are many successful applications in $\mathcal{N}=4$ SYM theory, one might also apply Janus deformations to the ABJM theory. However, the coupling constant of the $U(N)\times U(N)$ ABJM theory is $N/k$  with Chern-Simons level $k$, which is given by a discrete numerical value. So
it is unnatural to consider spatially varying coupling constant for the ABJM theory. For this reason, the Janus deformation of the coupling constant is not possible in the ABJM theory. On the other hand, this difficulty can be circumvented by introducing a parameter \cite{Honma:2008un} in the Bagger-Lambert-Gustavsson theory \cite{Bagger:2007jr}, which is another effective action of M2-branes. In \cite{Honma:2008un}, the authors considered Janus configurations using a spacetime dependent vacuum expectation value of a scalar field and a constant mass parameter.  
Although the Janus deformation of the ABJM theory has not been constructed yet, Janus solutions have already been studied in 11-dimensional supergravity and gauged $\mathcal{N}=8$ supergravity in 4-dimensions \cite{Estes:2012vm,DHoker:2008rje,Bobev:2013yra,Pilch:2015dwa}. In addition, other related Janus solutions were introduced in \cite{Karndumri:2016tpf,Gutperle:2017nwo,Suh:2018nmp}.

In this work, we construct a supersymmetric Janus mABJM model, that is, a model with a spacetime dependent mass term. The deformation preserves the half of supersymmetry of the mABJM theory. The mass parameter in the mABJM theory is originated from a constant 4-form flux \cite{Lambert:2009qw,Kim:2012gz}. The flux term is identified with the Wess-Zumino (WZ) type coupling. Our model admits a mass deformation after the supersymmetric completion for the flux term. 
As a result, we construct a large class of  $\mathcal{N}=3$ Janus mABJM models.

This paper is organized as follows. In section 2, we summarize the supersymmetry transformation of the mABJM theory and construct the Janus mABJM models with spacetime dependent mass parameters. In section 3, we clarify the origin of the spacetime-dependent mass term by using the WZ-type coupling. In section 4, we conclude with a summary and a discussion for future directions.

\section{${\cal N} = 3$ Supersymmetric Janus mABJM Models}
Like the Janus extension in the ${\cal N}=4$ super Yang-Mills theory~\cite{Clark:2004sb,DHoker:2006qeo,Kim:2008dj,Gaiotto:2008sd}, which was constructed by introducing the spacetime-dependent continuous coupling constant, one can introduce the spacetime dependence of the continuous mass parameter in the mABJM theory. In this section, we consider Janus extensions of the mABJM theory and construct the ${\cal N} =3$ supersymmetric Janus mABJM models with one spatial coordinate and one light-cone coordinate dependences, respectively. After reviewing the supersymmetric structure of the ${\cal N} =6$ mABJM theory with constant mass parameter, we construct the ${\cal N} =3$ supersymmetric Janus models in 3-dimensions.

\subsection{Supersymmetry of the ${\cal N} = 6$ mABJM theory}
We briefly summarize the supersymmetric structure of the ${\cal N} = 6$ ABJM theory with the action,
\begin{align}\label{ABJMac}
S =\int d^3x\,\left({\cal L}_0 + {\cal L}_{{\rm CS}} -V_{{\rm ferm}}
-V_{{\rm bos}} \right),
\end{align}
where 
\begin{align}
{\cal L}_0 &= {\rm tr}\left(-D_\mu Y_A^\dagger D^\mu Y^A +
i\psi^{\dagger A} \gamma^\mu D_\mu \psi_A\right),
\label{ABJMkin} \\
{\cal L}_{{\rm CS}} &= \frac{k}{4\pi}\,\epsilon^{\mu\nu\rho}\,{\rm tr}
\left(A_\mu \partial_\nu A_\rho +\frac{2i}{3}A_\mu A_\nu A_\rho
- \hat{A}_\mu \partial_\nu \hat{A}_\rho
-\frac{2i}{3}\hat{A}_\mu \hat{A}_\nu \hat{A}_\rho \right),
\label{ABJMcs} \\
V_{{\rm ferm}} &= \frac{2\pi i}k{\rm tr}\Big( Y_A^\dagger Y^A\psi^{\dagger
B}\psi_B -Y^A Y_A^\dagger\psi_B \psi^{\dagger B}
+2Y^AY_B^\dagger\psi_A\psi^{\dagger B} -2Y_A^\dagger
Y^B\psi^{\dagger A}\psi_B
\label{ferV} \\
&\hskip 1.7cm  +\epsilon^{ABCD}Y^\dagger_A\psi_BY^\dagger_C\psi_D
-\epsilon_{ABCD}Y^A\psi^{\dagger B}Y^C\psi^{\dagger D} \Big),
\nn \\
V_{{\rm bos}} &=-\frac{4\pi^2}{3k^2}{\rm tr}\Big(
Y^\dagger_AY^AY^\dagger_BY^BY^\dagger_CY^C
+Y^AY^\dagger_AY^BY^\dagger_BY^CY^\dagger_C
+4Y^\dagger_AY^BY^\dagger_CY^AY^\dagger_BY^C
\label{bosV} \\
&\hskip 2cm -6Y^AY^\dagger_BY^BY^\dagger_AY^CY^\dagger_C \Big).\nn
\end{align}
The supersymmetry transformation and convention of spinor representations are given in Appendix \ref{susyABJM}.

For later convenience to construct the Janus mABJM models, we divide the supersymmetric transformation rules in \eqref{N6part} as 
\begin{align}\label{delta1A2}
\delta = \delta_1 + \delta_A + \delta_2, 
\end{align}
where   
\begin{align}\label{delta1}
&\delta _1 Y^A =i \omega ^{\text{AB}} \psi _B, 
\qquad \delta _1 Y_A^{\dagger} = i \psi^{\dagger B}\omega_{AB}, 
\nn \\
&\delta_1\psi_A = \gamma^\mu \omega_{AB} D_\mu Y^B, 
\qquad 
\delta_1 \psi^{\dagger A} = -D_\mu Y_B^\dagger \omega^{AB}\gamma^\mu, 
\nn \\
&\delta_2 \psi_A= \frac{2\pi}{k} \omega_{AB}\left(Y^B Y_{C}^\dagger Y^C - Y^C Y_{C}^\dagger Y^B\right) + \frac{4\pi}{k} \omega_{BC} Y^B Y_A^\dagger Y^C,
\nn \\
&\delta_2 \psi^{\dagger A} = \frac{2\pi}{k} \omega^{AB}\left(Y_{C}^\dagger Y^C Y_B^\dagger- Y_{B}^\dagger Y^C Y_C^\dagger\right) - \frac{4\pi}{k} \omega^{BC} Y_B^\dagger Y^A Y_C^\dagger, 
\nn \\
&\delta_A A_\mu =  -\frac{2\pi}{k}\left(Y^A\psi^{\dagger B}\gamma_\mu\omega_{AB} + \omega^{AB} Y_A^\dagger  \gamma_\mu\psi_B \right), 
\nn  \\
&\delta_A \hat A_\mu = -\frac{2\pi}{k}\left(
\psi^{\dagger B} \gamma_\mu Y^A\omega_{AB} + \omega^{AB}Y_A^\dagger\gamma_\mu\psi_B \right).
\end{align}
Using these variations, we obtain the relations for the supersymmetric variations of terms in the Lagrangian \eqref{ABJMac} up to total derivatives,   
\begin{align}\label{delta1A}
&\delta_1 {\cal L}_0 + \delta_A {\cal L}_{{\rm CS}} = 0, 
\nn \\
&(\delta_A + \delta_2){\cal L}_0 = \delta_1 V_{{\rm ferm}}.
\end{align}
The remaining relation for supersymmetry variations, 
\begin{align}\label{delta21}
\delta_2 V_{{\rm ferm}} + \delta_1 V_{{\rm bos}} = 0, 
\end{align}
is automatically satisfied. Combining \eqref{delta1A} and \eqref{delta21}, 
we obtain 
\begin{align}\label{dnmass}
\delta \left( {\cal L}_0 + {\cal L}_{{\rm CS}} -V_{{\rm ferm}}
-V_{{\rm bos}}   \right)=0.
\end{align}

\subsection{Deformation by spacetime-dependent mass parameter}
We consider the spacetime dependent mass parameter, i.e., a Janus-type description of the ABJM theory with mass deformation. 
In addition to the supersymmetry transformation rules without mass deformation in \eqref{delta1}, we consider an additional supersymmetry transformation rules for fermions,
\begin{align}\label{JmSUSY}
&\delta_J \psi_A = m(x^\mu) M_A^{~B} \omega_{BC} Y^C, 
\nn \\
&\delta_J\psi^{\dagger A} = m(x^\mu) M^A_{~B} \omega^{BC} Y_C^\dagger,
\end{align}
where we introduce a diagonal mass matrix $M^A_{~B} = M_A^{~B} = {\rm diag} (1,1,-1,-1)$ and assume that the mass parameter $m$ depends on the worldvolume coordinates $x^\mu = (t,w_1,w_2)$ of M2-branes. 
Actually, the additional supersymmetry transformation rules in \eqref{JmSUSY} is exactly same with those of the mABJM theory~\cite{Hosomichi:2008jb,Gomis:2008vc}, expect for the spacetime dependence of the mass parameter $m$.

Acting the addition transformation rule \eqref{JmSUSY} to ${\cal L}_0$ in \eqref{ABJMkin}, we obtain the relation up to total derivatives,  
\begin{align}\label{dJL0}
\delta_J {\cal L}_0 -\delta_1 \hat V_{{\rm ferm}} = -i (\partial_\mu m) M^A_{~B} Y_C^\dagger \omega^{BC} \gamma^\mu \psi_A + i (\partial_\mu m) M_{A}^{~B}\psi^{\dagger A}\gamma^\mu \omega_{BC} Y^C, 
\end{align}
where 
\begin{align}\label{Vhferm}
\hat V_{{\rm ferm}} = i m M_{A}^{~B} \psi^{\dagger A}\psi_B.
\end{align}
Here and from now on we omit `tr' for simplicity. 
We also obtain the relation 
\begin{align}\label{dmVf}
\delta_J V_{{\rm ferm}} + (\delta_2 + \delta_J) \hat V_{{\rm ferm}} + \delta_1 \hat V_{{\rm bos}} = 0,
\end{align}
where $V_{{\rm ferm}}$ is given in \eqref{ferV} 
and we introduce the deformed bosonic potential term, 
\begin{align}\label{bosdef}
\hat V_{{\rm bos}} = \hat V_{{\rm flux}} + \hat V_{{\rm mass}}
\end{align}
with 
\begin{align}\label{flma}
\hat V_{{\rm flux}} = -\frac{4\pi m}{k} M_B^{~D}\left(Y_C^\dagger Y^C Y_D^\dagger Y^B - Y^C Y_C^\dagger Y^B Y_D^\dagger\right),\quad
\hat V_{{\rm mass}} = m^2 Y_A^\dagger Y^A.
\end{align}
Combining \eqref{dnmass}, \eqref{dJL0}, and \eqref{dmVf},  we obtain 
\begin{align}\label{tdelta}
\delta_{{\rm tot}} {\cal L}_{\rm mABJM}
= -i (\partial_\mu m) M^A_{~B} Y_C^\dagger \omega^{BC} \gamma^\mu \psi_A + i (\partial_\mu m) M_{A}^{~B}\psi^{\dagger A}\gamma^\mu \omega_{BC} Y^C,
\end{align}
where $\delta_{{\rm tot}} = \delta + \delta_J$ with definitions in \eqref{delta1A2} and \eqref{JmSUSY} and 
${\cal L}_{\rm mABJM}$ is the Lagrangian of the mABJM theory~\cite{Hosomichi:2008jb,Gomis:2008vc}, 
\begin{align}\label{mABJM}
{\cal L}_{\rm mABJM} = {\cal L}_0 +{\cal L}_{{\rm CS}} 
- V_{{\rm ferm}} - V_{{\rm bos}}-\hat V_{{\rm ferm}}   -\hat V_{{\rm bos}}. 
\end{align}
As we see in \eqref{tdelta}, the Lagrangian \eqref{mABJM} of the mABJM theory is not invariant under the total  supersymmetry transformation $\delta_{tot}$ for the case $\partial_\mu m\ne 0$. Therefore, in order to cancel the right-hand-side of \eqref{tdelta}, one has to impose some constraints to the supersymmetric parameters, which reduce the number of the supersymmetry, and add some terms in the Lagrangian.   In this work, we break half of the full supersymmetry by imposing a constraint on the supersymmetric parameter $\omega^{AB}$ and add some bosonic potential terms to cancel the right-hand-side of \eqref{tdelta}.  

\subsubsection{${\cal N}=3$ Janus model with $m=m(w_1)$}
In the case that the mass parameter has one spatial direction dependency, $m= m(w_1)$, 
an ${\cal N}=3$ supersymmetric Janus model can be constructed by adding some interaction term to the mABJM Lagrangian in \eqref{mABJM}. The procedure to construct the Janus model is following. 
The right-hand-side of \eqref{tdelta} with the spatial dependence $m= m(w_1)$ is reduced to 
\begin{align}\label{RHSmx}
{\rm RHS} = -i m' M^A_{~B} Y_C^\dagger \omega^{BC} \gamma^1 \psi_A + i m' M_{A}^{~B}\psi^{\dagger A}\gamma^1 \omega_{BC} Y^C,
\end{align}
where $m'\equiv \partial m/\partial w_1$.
Therefore, the Lagrangian of the mABJM theory with $m'\ne 0$ is not invariant under the full ${\cal N}=6$ supersymmetry transformation. 
Now we investigate whether there exist some lower supersymmetric models under the transformations, $\delta$ and $\delta_J$ defined in \eqref{delta1} and \eqref{JmSUSY}, respectively.   
To do that, we impose some restrictions on the supersymmetric parameter $\omega^{AB}$. For the detailed properties of $\omega^{AB}$, see Appendix \ref{susyABJM}.

According to the structure of supersymmetric multiplets $(Y^A,\psi^{\dagger A})$ of the ABJM theory, one can divide the supersymmetric parameter $\omega^{AB}$ into $\omega^{ab}$ (or $\omega^{ij}$) and $\omega^{ai}$, where $a,b=1,2$, and $i,j = 3,4$. Here $\omega^{ab}$ and $\omega^{ij}$ are connected by the reality condition \eqref{omegaAB}.  $\omega^{ab}$ and $\omega^{ai}$ parametrize the 4 (${\cal N} =2$) and 8 (${\cal N} =4$) supercharges, respectively.  
Now we consider the following projections for $\omega^{AB}$ in terms of the gamma matrix $\gamma^1$,  
\begin{align}\label{proj}
\gamma^1 \omega_{ab}=-\omega_{ab}\quad &\Longleftrightarrow\quad \omega^{ab}\gamma^1 =\omega^{ab}, 
\nn \\
\gamma^1 \omega_{ai}=\omega_{ai}\quad &\Longleftrightarrow\quad \omega^{ai}\gamma^1 =-\omega^{ai},
\end{align}
where we used the spinor conventions in \eqref{xichi} and \eqref{gampro}\footnote{The projections in \eqref{proj} in terms of explicit spinor indices are rewritten as 
\begin{align}
&\gamma_\alpha^{1~\beta} \omega_{ab\beta}= - \omega_{ab\alpha},
\nn \\
& \omega^{ab\beta}\gamma^{1~\alpha}_{\beta}= \gamma^{1\alpha}_{~~\beta}\omega^{ab\beta} = -\gamma^{1\alpha\beta}\omega^{ab}_{~~\beta} = \omega^{ab\alpha},
\end{align} 
where we choose $-1$ as the eigenvalue of $\gamma^1 = \sigma^1$. 
The same procedure can be applied for $\omega_{ai}$ by choosing $+1$ as the eigenvalue of $\gamma^1$ in \eqref{proj}. }.
The projections in \eqref{proj} break half of the ${\cal N} = 6$ supersymmetry represented by the parameter $\omega^{AB}$ and preserve the ${\cal N} = 3$ supersymmetry. Expanding the RHS in \eqref{RHSmx} and applying the projection in \eqref{proj}, we obtain 
\begin{align}\label{approj}
{\rm RHS} = &\,i m'\left( Y_b^\dagger\omega^{ba}\psi_a - Y_i^\dagger\omega^{ia}\psi_a +Y_a^\dagger\omega^{ai}\psi_i - Y_j^\dagger\omega^{ji}\psi_i\right)
\nn \\
-&i m'\left(-\psi^{\dagger a}\omega_{ba} Y^b + \psi^{\dagger a}\omega_{ia} Y^i - \psi^{\dagger i}\omega_{ai} Y^a + \psi^{\dagger i}\omega_{ji} Y^j\right).
\end{align}
Decomposing the supersymmetry transformation rule of the scalar fields in \eqref{delta1}, we have the relations,
\begin{align}\label{dYdecom}
\delta Y^a = i\omega^{ab}\psi_b + i\omega^{ai}\psi_i,\quad&\Longleftrightarrow\quad \delta Y_a^\dagger = i \psi^{\dagger b}\omega_{ab} + i\psi^{\dagger i}\omega_{ai},
\nn \\
\delta Y^i = i\omega^{ia}\psi_a + i\omega^{ij}\psi_j,\quad&\Longleftrightarrow\quad \delta Y_i^\dagger = i \psi^{\dagger a}\omega_{ia} + i\psi^{\dagger j}\omega_{ij},
\end{align}
where we replaced $\delta_1$ to $\delta$ for scalar fields. Using the relations in \eqref{dYdecom}, we rewrite the RHS in \eqref{approj} as 
\begin{align}\label{RHS2}
{\rm RHS} &= m' \left(Y_a^\dagger\delta Y^a - Y_i^\dagger \delta Y^i + \delta Y_a^\dagger Y^a-\delta Y_i^\dagger  Y^i\right)
\nn \\
&= m' \delta\left(Y_a^\dagger Y^a- Y_i^\dagger  Y^i\right)
= m' \delta\left(M_A^{~B} Y_B^\dagger Y^A\right).
\end{align}

Combining \eqref{tdelta}, \eqref{RHSmx}, and \eqref{RHS2},  
we obtain 
\begin{align}\nn
\delta_{{\rm tot}}\left({\cal L}_{\rm mABJM} - V_J\right)=0
\end{align}
with 
\begin{align} \label{VJ}
V_J = m' M_A^{~B} Y_B^\dagger Y^A,
\end{align} 
where we replace $\delta$ to $\delta_{{\rm tot}}$ since $\delta_J V_J = 0$. 
Then the resulting action 
\begin{align}\label{Janusac}
S_{{\rm Janus}} = \int d^3 x \left({\cal L}_{\rm mABJM} - V_J\right)
\end{align}
has ${\cal N} = 3$ supersymmetry.

\subsubsection{${\cal N}=3$ Janus model with $m=m(t\pm w_1)$}
We also find that the mABJM theory in \eqref{mABJM} has the ${\cal N}=3$ supersymmetry in the case of the light-cone coordinate dependence of the mass parameter, $m = m(t\pm w_1)$. In this case, the right-hand-side of \eqref{tdelta} is reduced to 
\begin{align}\label{ltcone}
{\rm RHS} = -i m' M^A_{~B} Y_C^\dagger \omega^{BC} \left(\gamma^0 \pm \gamma^1\right) \psi_A + i m' M_{A}^{~B}\psi^{\dagger A}\left(\gamma^0 \pm \gamma^1\right) \omega_{BC} Y^C,
\end{align}
where $m'\equiv  \partial m/\partial t = \partial m/\partial w_1$. 
Since $\gamma^0 = i\sigma^2$ and $\gamma^1=\sigma^1$, 
one can consider the following projection 
\begin{align}\label{ltproj}
\gamma_\pm \omega_{AB} = 0, 
\end{align}
where $\gamma_\pm \equiv \gamma^0 \pm \gamma^1$. In this projection, the RHS in \eqref{ltcone} is vanishing and the ${\cal N}=6$ supersymmetry of the supersymmetric parameter $\omega_{AB}$ is reduced to ${\cal N}=3$ by the projection.   
That is, the Lagrangian of the mABJM theory in \eqref{mABJM} itself has the ${\cal N}= 3$ supersymmetry for the lightcone coordinate dependence of the mass parameter.

\section{M2-branes on the Space-dependent Background Flux}
In the previous section, we considered the ${\cal N}=3$ supersymmetric completion in the presence of the 
spacetime-dependent mass parameter in the field theory point of view. In this section, we consider the physical origin for the case $m= m(w_1)$ in M-theory. 
The dynamics of the M2-branes in the presence of the background 3-form gauge fields in 11-dimensional supergravity is described by WZ-type coupling in the ABJM theory. The phenomenon, such as the Myers effect in string theory~\cite{Myers:1999ps}, occurs in the M-theory, and so the interaction with the 6-form gauge field is also described by the WZ-type coupling. 
In order to introduce such WZ-type couplings as  interaction terms in the M2-brane theory, the non-Abelian U($N)\times{\rm U}(N)$ gauge symmetry of the ABJM theory should be preserved by the interaction terms. 
In relation with this issue, there were several works in the literature \cite{Li:2008ez,Ganjali:2009kt,Kim:2009nc,Lambert:2009qw,Sasaki:2009ij,Kim:2010hj,Allen:2011pm,Kim:2012gz,Jang:2015efa}. Specially in \cite{Kim:2009nc,Kim:2010hj,Allen:2011pm,Kim:2012gz,Jang:2015efa}, the gauge invariant WZ-type couplings were constructed under the assumption that the 3-form and 6-form gauge fields depend on scalar fields $Y^A,\, Y_A^\dagger$ as well as worldvolume coordinates $(t,w_1,w_2)$ of the M2-branes. 
The deformed quartic potential term $\hat V_{{\rm flux}}$ in \eqref{flma} in the mABJM theory can be identified with the WZ-type coupling for the 6-form gauge field in the presence of the constant self-dual 4-form field strength~\cite{Lambert:2009qw,Kim:2010hj, Kim:2012gz}. For the supersymmetric completion in the presence of the constant 4-form field strength, one needs the mass term in \eqref{flma} as well. The origin of the mass term in the M-theory point of view was investigated in \cite{Lambert:2009qw}. In order to discuss the physical origin of the space dependent mass parameter in the ${\cal N} = 3$ supersymmetric action \eqref{Janusac}, we mainly follow the discussion of \cite{Lambert:2009qw} and \cite{Kim:2012gz}.

We consider the 4-form field strength in 11-dimensional supergravity, which has one spatial coordinate dependence in the world-volume of M2-branes, 
\begin{align}\label{TABCD0}
F_{AB\bar C\bar D} =  T_{AB\bar C\bar D}(w_1), 
\end{align}
where the complex-valued parameters $T_{AB\bar C\bar D}= (T_{CD\bar A\bar B})^*$ are anti-symmetric in two indices $C,D$ as well as $A,B$. Here we use four complex coordinates $y^A$ $(A=1,2,3,4)$ in the transverse direction to the M2-brane worldvolume having the relation with the eight transverse coordinates $x^I$ $(I=1,\cdots 8)$, 
\begin{align}
y^A = x^A + i x^{A+4}. 
\end{align}
The corresponding (anti)-bifundamental fields for the complexified coordinate $y^A$ and $y^{A\dagger}$ are $Y^A$ and $Y_A^\dagger$, respectively. We also employed the index notations, where unbarred indices are contracted with bifundamental fields, while barred ones are contracted with anti-bifundamental fields. For the details of notations, see \cite{Kim:2010hj}. 
Under a special choice for the constant 4-form tensor~\cite{Kim:2012gz}, 
\begin{align}\label{TT}
T_{12\bar 1\bar 2} = -m,\quad T_{34\bar 3\bar 4} = m,
\end{align}
the WZ-type coupling for the 6-form gauge field is reduced to the flux term \eqref{flma} in the infinite tension limit of M2-branes~\cite{Lambert:2009qw} in the ${\cal N}=6$ mABJM theory.

We try to construct the WZ-type coupling corresponding to the necessary 4-form field strength configuration to get the space-dependent background \eqref{TABCD0} in 11-dimensional supergravity.  Since the 4-form field strength $F_4$ and the 3-form gauge field $C_3$ have the relation $F_4 = dC_3$ in 11-dimensional supergravity,  one can rewrite the equation $dF_4=0$ for the case \eqref{TABCD0} as 
\begin{align}\label{parFAB}
\partial_{[w_1} F_{AB\bar C\bar D]} = 0. 
\end{align}
Here we set other components of 4-form field strengths, which do not appear in \eqref{parFAB}, to zero by using a suitable gauge transformation $\delta C_3 = d\Lambda_2$. 
This relation is expanded as 
\begin{align}\label{FxBCD}
\partial_{w_1} F_{AB\bar C\bar D} - \partial_A F_{w_1 B\bar C\bar D} + \partial_B F_{w_1A\bar C\bar D} - \partial_{\bar C} F_{ABw_1\bar D} + \partial_{\bar D} F_{ABw_1\bar C} = 0, 
\end{align}
where $(F_{w_1 B\bar C\bar D})^* = F_{CDw_1\bar B}$, $\partial_A\equiv \partial/\partial (\lambda Y^A)$, $\partial_{\bar A}\equiv \partial/\partial (\lambda Y_A^\dagger)$, and $\lambda = 2\pi l_{{\rm P}}^{3/2}$ with Planck length $l_{{\rm P}}$\footnote{Since the mass dimension of the scalar field $Y^A$ is $\frac12$ in the ABJM theory, the derivative operator $\partial/\partial Y^A$ has mass dimension $-\frac12$. In order to unify the mass dimensions of derivative operators  including the operator $\partial_{w_1}$ in \eqref{FxBCD}, we rescale the derivative operators in terms of Planck length in M-theory. }. From \eqref{FxBCD}, we determine the form of $F_{w_1B\bar C\bar D}$ in terms of the transverse scalar fields\footnote{In the construction of the WZ-type coupling, the dependence of form fields on   the transverse scalars seems natural~\cite{Kim:2009nc,Kim:2010hj,Allen:2011pm,Kim:2012gz,Jang:2015efa}. Therefore, the field strengths also depend on the transverse scalars.}, 
\begin{align}\label{FxBCD2}
F_{w_1B\bar C\bar D} = \frac{\lambda}4 T'_{AB\bar C\bar D}(w_1) Y^A, \qquad 
F_{ABw_1\bar D} = \frac{\lambda}4 T'_{AB\bar C\bar D}(w_1) Y_C^\dagger,
\end{align}
where we consider the case that $T_{AB\bar C\bar D}$ is a real quantity and define $T'_{AB\bar C\bar D} \equiv \partial T_{AB\bar C\bar D}/\partial w_1$.

In order to figure out possible WZ-type couplings corresponding to the field strength configuration \eqref{FxBCD2}, we start from the general WZ-type couplings for $C_3$ and $C_6$~\cite{Kim:2012gz} in the ABJM theory, 
\begin{align}
S_{{\rm WZ}} = S_{C}^{(3)}+ S_{C}^{(6)}
\end{align} with
\begin{align}\label{act3}
S_{C}^{(3)} &= \mu_2\int d^3x\, \frac{1}{3!}
\epsilon^{\mu\nu\rho}\left\{{\rm Tr}\right\}\Big[ C_{\mu\nu\rho} +
3\lambda C_{\mu\nu A} D_\rho Y^A + 3\lambda^2 \big(C_{\mu AB} D_\nu
Y^A D_\rho Y^B + C_{\mu A\bar B} D_\nu Y^A D_\rho Y_B^\dagger \big)
\nonumber \\
&~~~~~~~~~~+ \lambda^3 (C_{ABC} D_\mu Y^A D_\nu Y^B D_\rho Y^C + C_{AB\bar C}
D_\mu Y^A D_\nu Y^B D_\rho Y_C^\dagger\big) + ({\rm c.c.}) \Big], 
 \\
S_{C}^{(6)} &= -\frac{\pi\lambda}{k}\mu_2\int d^3x\, \frac{1}{3!}
\epsilon^{\mu\nu\rho} \left\{{\rm Tr}\right\}\Big( C_{\mu\nu\rho A
B\bar C} \beta^{AB}_{~C}
+3\lambda \big(C_{\mu\nu ABC\bar D} D_\rho Y^A \beta^{BC}_{~D}
+ C_{\mu\nu AB\bar C\bar D} D_\rho Y_C^\dagger \beta^{AB}_{~D}
\big)
\nonumber \\
&+3\lambda^2 \big(C_{\mu ABCD\bar E} D_\nu Y^A D_\rho Y^B \beta^{CD}_{~E}
+C_{\mu ABC\bar D\bar E} D_\nu Y^A D_\rho Y_D^\dagger \beta^{BC}_{~E}
+C_{\mu AB\bar C\bar D\bar E} D_\nu Y_C^\dagger D_\rho
Y_D^\dagger \beta^{AB}_{~E} \big)
\nonumber \\
&+\lambda^3\big(C_{ABCDE\bar F}D_\mu Y^A D_\nu Y^B D_\rho Y^C \beta^{DE}_{~F}
+C_{ABCD\bar E\bar F}D_\mu Y^A D_\nu Y^B D_\rho Y_E^\dagger
\beta^{CD}_{~F}
\nonumber \\
&+C_{ABC\bar D\bar E\bar F}D_\mu Y^A D_\nu Y_D^\dagger D_\rho
Y_E^\dagger \beta^{BC}_{~F}
+C_{AB\bar C\bar D\bar E\bar F}D_\mu Y_C^\dagger D_\nu
Y_D^\dagger D_\rho Y_E^\dagger \beta^{AB}_{~F}\big) + ({\rm
c}.{\rm c}.) \Big).  \label{act6}
\end{align}
Here $\mu_2$ denotes the tension of M2-brane, which is proportional to $1/\lambda^2$,   $\{{\rm Tr}\}$ represents all  possible contractions of gauge indices among the form fields and transverse scalars to give single traces only and $\beta^{AB}_{~C}\equiv\frac12(Y^AY_C^\dagger Y^B - Y^BY_C^\dagger Y^A)$. 
Keeping in mind that the 6-form gauge field $C_6$ has the relation with the 4-form field strength as $dC_6 = *F_4 + \frac12 C_3\wedge F_4$, one can determine the scalar field dependence of $C_3$ and $C_6$ in terms of the information of $F_4$. In the presence of the background $F_4$ in \eqref{TABCD0} and \eqref{FxBCD2}, the gauge fields $C_3$ and $C_6$ have the scalar field dependence as 
\begin{align}\label{nvC}
&C_{w_1A\bar B} \sim \lambda^2T'_{AC\bar B\bar D} Y^CY_D^\dagger, \nn \\
&C_{AB\bar C} \sim \lambda T_{AB\bar C\bar D} Y_D^\dagger, 
\nn \\
& C_{\mu\nu\rho A B\bar C},~C_{\mu\nu\rho A \bar B\bar C},\cdots\, {\rm are~ linear~ in~ transverse~scalars,}
\nn \\
&C_{\mu\nu A B C D},~ C_{\mu\nu A B C\bar D},\cdots\, {\rm are~ linear~ or~ quadratic~ in~ transverse~scalars,}
\nn \\
&C_{w_1 A B C D E},~ C_{w_1 A B C D \bar E},\cdots\, {\rm are~ up~ to~ cubic~ order~in~ transverse~scalars,}
\nn \\
&C_{ABCDE\bar F},~C_{ABCD\bar E\bar F}\cdots {\rm are~ linear~ or~ quadratic~ in~ transverse~scalars}.
\end{align}
All non-vanishing components of form fields include $T_{AB\bar C \bar D}$ or $T'_{AB\bar C\bar D}$ factors and scalar fields. Other components can be set to zero using the U(1) gauge transformation of $C_3$ and $C_6$.

In this paper, we consider the WZ-type coupling in the infinite tension limit of the M2-branes ($\lambda\to 0$) to turn off couplings to gravity modes~\cite{Lambert:2009qw}. 
Due to the behavior of form fields in \eqref{nvC}, the WZ-type couplings in \eqref{act3} and \eqref{act6}  are vanishing in the limit $\lambda\to 0$, except for the following WZ-type couplings for 6-form gauge fields,  
\begin{align}\label{TABCD}
S_{\rm WZ} &= -\frac{\pi}{\lambda k}\int d^3x\, \frac{1}{3!} \epsilon^{\mu\nu\rho} {\rm tr}\big[ C_{\mu\nu\rho A B\bar C} \beta^{AB}_{~C} + C^{\dagger}_{\mu\nu\rho AB\bar C} (\beta_{~C}^{AB})^\dagger \big]
\nn \\
&=\frac{4\pi}{k}\int d^3x\,{\rm tr}\big(T_{AB\bar C\bar D}Y_C^\dagger Y^AY_D^\dagger Y^B \big),
\end{align}
where we set $\mu_2 = 1/\lambda^2$, and  
$C_{\mu\nu\rho A B\bar C} =- 2\lambda
\epsilon_{\mu\nu\rho} T_{AB\bar C\bar D}Y_D^\dagger,\,\,
C^\dagger_{\mu\nu\rho A B\bar C} =-2\lambda
\epsilon_{\mu\nu\rho} T_{CD\bar A\bar B}Y^D$.
The normalization of parameters can be adjusted by $T_{AB\bar C\bar D}$. Inserting \eqref{flma} and \eqref{TT} into \eqref{TABCD}, we obtain \begin{align}\label{SWZ}
S_{{\rm WZ}} = -\int d^3 x\, \hat V_{{\rm flux}}.
\end{align}
Therefore, we notice that the flux term  in the mABJM theory is identified with the WZ-type coupling for the self-dual 4-form field strength, which varies along one spatial coordinate of the worldvolume of the M2-branes in $\lambda\to0$ limit. As we see in \eqref{SWZ}, the flux term of the Janus mABJM model is the same with that of the mABJM theory, though the mass parameter is not constant, i.e $m = m(w_1)$,  in the current case.  
This argument for the physical origin of the flux term in the ${\cal N}=3$ Janus ABJM model with space-dependent mass parameter is also applicable to the case of ${\cal N}=3$ model with the light-cone coordinate dependence of the mass parameter.

For the given flux term in \eqref{TABCD}, one can implement the supersymmetric completion with the  parameter $\omega^{AB}$ under the constraints in \eqref{proj}, which generates the deformed mass term in \eqref{Janusac} and defines the ${\cal N} =3$ supersymmetric Janus model. It was also argued that the physical origin of the mass-squared term in \eqref{flma} is the metric response satisfying the Einstein equation in the presence of the nontrivial flux with constant mass parameter in \eqref{TT}~\cite{Lambert:2009qw}.
According to the same procedure with \cite{Lambert:2009qw}, one may obtain the deformed mass term with varying mass parameter in \eqref{Janusac} from the metric response in the presence of the flux term \eqref{TABCD}. 
Investing the metric response due to the flux term in the Janus model is intriguing but will not be pursed in this work.

\section{Conclusion}
In order to construct the ${\cal N} = 6$ mABJM theory, one has to introduce the deformed supersymmetry transformation \eqref{JmSUSY} with a constant mass parameter $m$. Then the ${\cal N} =6$ supersymmetric completion allows the fermionic mass term in \eqref{Vhferm} and the bosonic potential term in \eqref{bosdef}, which are composed of the flux and mass terms, respectively. 
In this paper, we considered two kinds of spacetime-dependent mass parameters, $m= m(w_1)$ and $m = m(t\pm w_1)$, with the same deformed supersymmetry transformation rule. We found the maximal supersymmetry for these cases is ${\cal N}=3$ with suitable constraints to the supersymmetry parameter $\omega^{AB}$. For the $m = m(w_1)$ case, we showed that the mass term $\hat V_{{\rm mass}}$ in \eqref{flma} should be deformed to 
\begin{align}\label{Jmass}
\left(m^2\delta^B_A + m' M_A^{~B}\right) Y^AY_B^\dagger = \left(m^2 + m'\right) Y^a Y_a^\dagger + \left(m^2 - m'\right) Y^i Y_i^\dagger ~,
\end{align}
where $a = 1,2$ and $i = 3,4$, and the supersymmetry parameter should satisfy \eqref{proj}.
On the other hand, for the $m = m(t\pm w_1)$ case, we found that the ${\cal N} =3$ Janus Lagrangian is same with that of the mABJM theory. However, the supersymmetry parameter $\omega^{AB}$ has to satisfy the constraints \eqref{ltproj}.

We also discussed the origin of the Janus mass parameter in M-theory. The mass parameter of the mABJM theory is identified with the constant self-dual 4-form field strength in the transverse space of the M2-branes. Under the constant background field strength, the corresponding WZ-type coupling in the infinite tension limit of M2-branes is the same with the flux term in \eqref{flma}. Along the same line of the above discussion, we considered the spacetime dependent 4-form field strength and showed that the corresponding WZ-type coupling is also identical to the flux term.

In quantum field theory point of view, our models are ${\cal N} = 3$ Cherm-Simons matter theories with spacetime-dependent mass gap. It would be interesting to investigate the characteristics of those models, such as the vacuum structure, soliton solutions, and an extension to lower supersymmetric Janus models. Furthermore, specific models can be applied to condensed matter physics by introducing various spacetime-dependent mass functions.

It was known that the mABJM theory is dual to the 11-dimensional supergravity on the LLM solutions~\cite{Lin:2004nb}. In this duality, the mass parameter is related to a self-dual 4-form field strength. One of the crucial differences between the mABJM theory and our Janus models is the spacetime-dependence of mass parameter. For this reason, one can also expect that there exist gravity duals of our Janus models, as less supersymmetric solutions in 11-dimensional supergravity. We leave this issue as a future work~\cite{ShinOK}.

\section*{\bf Acknowledgement}

K.Kim appreciates APCTP for its hospitality during completion of this work.  This work was supported by National Research Foundation of Korea(NRF) grant with grant numbers NRF-2017R1D1A1A09000951(O.K.) and NRF-2015R1D1A1A01058220(K.K.). 

\vspace{0cm}

\appendix 
\section{Supersymmetric Structure of the ${\cal N}=6$ ABJM Theory}\label{susyABJM}
The action \eqref{ABJMac} is invariant under the ${\cal N}=6$ supersymmetry transformation, 
\begin{align}
&\delta Y^A = i \omega^{AB}\psi_B,
\nn \\
&\delta Y^{\dagger}_A = i \psi^{\dagger B}\omega_{AB},
\nn \\
&\delta\psi_A = \gamma^\mu\omega_{AB}D_\mu Y^B
+\frac{2\pi}k\omega_{AB}(Y^BY^\dagger_CY^C -Y^CY^\dagger_CY^B)
+\frac{4\pi}k\omega_{BC}Y^BY^\dagger_AY^C,
\nn \\
&\delta\psi^{\dagger\,A}= -D_\mu Y^\dagger_B \omega^{AB}\gamma^\mu
+\frac{2\pi}k\omega^{AB}(Y^\dagger_CY^CY^\dagger_B
-Y^\dagger_BY^CY^\dagger_C)
-\frac{4\pi}k\omega^{BC}Y^\dagger_BY^AY^\dagger_C,
\nn \\
&\delta A_\mu =  -\frac{2\pi}{k}\left(Y^A\psi^{\dagger B}\gamma_\mu\omega_{AB} + \omega^{AB} Y_A^\dagger  \gamma_\mu\psi_B \right), 
\nn  \\
&\delta \hat A_\mu = -\frac{2\pi}{k}\left(
\psi^{\dagger B} \gamma_\mu Y^A\omega_{AB} + \omega^{AB}Y_A^\dagger\gamma_\mu\psi_B \right),
\label{N6part}
\end{align}
where the supersymmetric parameter $\omega^{AB}$ satisfies the reality condition,  
\begin{align}\label{omegaAB}
\omega^{AB}=-\omega^{BA}=(\omega_{AB})^*=
\frac{1}{2}\,\epsilon^{ABCD}\omega_{CD}.
\end{align}
More specifically, we impose the relations for the supersymmetric parameters, 
\begin{align}
&\omega_{AB} = \epsilon_i(\Gamma^i)_{AB}, \quad \omega^{AB} = (\omega_{AB})^* = \epsilon_i(\Gamma^{i*})^{AB}, 
\nn \\
& \frac12 \epsilon^{ABCD}\omega_{CD} = \frac12 \epsilon^{ABCD}\epsilon_i(\Gamma^i)_{CD} = -\epsilon_i (\Gamma^{i\dagger})^{AB}= -\epsilon_i (\Gamma^{i*})^{BA} = \omega^{AB},
\end{align}
where $i= 1,2,\cdots 6$, $\epsilon_i$'s represent (2+1)-dimensional Majorana spinors, $\Gamma^i$'s are 4$\times$4 matrices, called chirally decomposed 6-dimensional $\Gamma$-matrices satisfying the conditions, 
\begin{align}
\{\Gamma^i,\, \Gamma^{j\dagger}\} = 2\delta^{ij},\qquad 
(\Gamma^i)_{AB} = - (\Gamma^i)_{BA}. 
\end{align} 
One example of $\Gamma^i$-matrices is 
\begin{align}
&\Gamma^1 = \sigma^2\otimes \mathbb{I}_2,\quad 
\Gamma^2 = -i\sigma^2\otimes \sigma^3,
\nn \\
&\Gamma^3 = i\sigma^2\otimes \sigma^1,\quad 
\Gamma^4 = -\sigma^1\otimes \sigma^2,
\nn \\
&\Gamma^5 = \sigma^3\otimes \sigma^2,\quad 
\Gamma^6 = -i\mathbb{I}_2\otimes \sigma^2.
\end{align}

Here, we used the convention of spinor indices for the two component spinors $\xi$ and $\chi$,
\begin{align}\label{xichi}
&\xi^\alpha = \epsilon^{\alpha\beta}\xi_\beta,\quad \xi_\alpha = \epsilon_{\alpha\beta}\xi^\beta,\quad\xi\chi \equiv\xi^\alpha\chi_\alpha, \quad\xi\gamma^\mu\chi\equiv \xi^\alpha\gamma_\alpha^{\mu\beta}\chi_\beta,
\end{align}  
where $\alpha,\,\beta=1,2$ represent spinor indices and $\epsilon^{12}=-\epsilon^{21}=-\epsilon_{12}=\epsilon_{21}=1$. The 3-dimensional Gamma matrices have the form, 
\begin{align}
\gamma^\mu = (i\sigma^2, \sigma^1,\sigma^3)
\end{align}
with the properties, 
\begin{align}\label{gampro}
&\gamma^\mu\gamma^\nu = \eta^{\mu\nu} + \epsilon^{\mu\nu\rho}\gamma_\rho, 
\quad\gamma^{\mu~\beta}_{\alpha} \gamma^{~~\beta'}_{\mu\alpha'} = 2\delta_{\alpha'}^{\beta}\delta_{\alpha}^{\beta'} - \delta_{\alpha}^{\beta}\delta_{\alpha'}^{\beta'},
\quad\gamma_\alpha^{\mu~\beta} = \gamma_{~~~\alpha}^{\mu\beta}. 
\end{align}

\end{document}